\documentclass[twocolumn,showpacs,preprintnumbers,amsmath,amssymb,prb,aps]{revtex4-1}
\usepackage{epsfig}
\usepackage{epstopdf}
\usepackage{multirow}
\usepackage{graphicx}
\usepackage{dcolumn}
\usepackage{bm}
\usepackage{textcomp}
\usepackage{array}
\usepackage{csquotes}
\usepackage{subcaption}
\usepackage{booktabs}
\usepackage{amsmath}
\usepackage{hyperref}
\hypersetup{
    colorlinks=true,
    linkcolor=blue,
    filecolor=magenta,      
    urlcolor=blue,
    pdftitle={},
    }
    
\begin{document}
\title{ Lattice dynamics related properties of Nickel: A comparative DFT and DFT+U study}

\author{Shivani Bhardwaj$^{1,}$}
\altaffiliation{Electronic mail: spacetimeuniverse369@gmail.com}

\author{Sudhir K. Pandey$^{2,}$}
\altaffiliation{Electronic mail: sudhir@iitmandi.ac.in}
\affiliation{$^{1}$School of Basic Sciences, Indian Institute of Technology Mandi, Kamand - 175075, India}
\affiliation{$^{2}$School of Mechanical and Materials Engineering, Indian Institute of Technology Mandi, Kamand - 175075, India}


\date{\today} 

\begin{abstract}
The simultaneous influence of electronic correlations and magnetic ordering on the theoretical estimation of phonons and related properties of Ni is investigated. The work includes a comparative DFT and DFT+U study, where on-site Coulomb interaction parameter for 3$d$ electrons, $U$($U_{full}$)= 0.516 eV obatined from constarined random phase approximation (cRPA) calculations, is considered for DFT+U calculations. The analysis of phonon frequency estimates along high symmetric k-directions and sampled full-BZ (Brillouin zone) using Frozen phonon displacement method suggests the importance of both on-site Coulomb correlations and magnetism to account for the experimental frequencies. Further, prominent role of both the aspects is observed in the derived thermodynamic properties - Free-energy, specific heat \& entropy, within quasi-harmonic approximation (QHA) specially at high temperatures. The temperature dependent evaluation of thermal expansion coefficient($\alpha$) and phonon density of states is performed together with the equilibrium elastic constants. The results obtained for Ni, suggest the significance of electronic energy correction due to both on-site Coulomb correlations and magnetic phase incorporation, to account for realistic description of experimental findings.
This study realizes the inevitable role of correlation effects in studying the phononic properties of a correlated transition metal, hence directing a way to explore various other correlated electron systems for their lattice dynamics.

\vspace{0.2cm}
\end{abstract}

\maketitle

\section{Introduction}

Lattice dynamics of a solid is said to be mainly dictated by its ionic degrees of freedom. However, on theoretical grounds, the consideration of electronic degrees of freedom is found to have significant impact on the estimation of lattice dynamic properties along with spin-degrees of freedom, especially in systems with magnetic ordering \cite{kim}. Evidently, to an extent such considerations have been investigated for their cruciality in correlated electron systems where profound interplay between correlation effects and magnetism could be expected to directly affect their lattice dynamic properties. Essentially, in numerical respects, the determination of phononic and related thermal properties which in essence are derived from electronic energy can be well suspected to have primary dependence on the modification in electronic energy inhabiting correlation effects' correction \cite{Martin,prakash}.  

One such work by \textit{Corso} 
 $et. al.$ \cite{PhysRevB.62.273} on Fe and Ni to study the effect of magnetization on phonon frequencies conclude by regarding the effect of magnetization to be quite small, suggesting the implementation of combined use of ultrasoft pseudopotentials, spin polarized generalized gradient approximations (GGA) and non-linear core corrections within Density functional perturbation theory (DFPT) technique to account for the experimental frequencies. On the contrary, \textit{Lee} $et. al.$ \cite{Lee} show explicit dependence of phonon frequencies in Ni, on its magnetic moment within density-functional-based linear-response framework.
 The effect of magnetization on thermal properties has also been reported by \textit{Hatt} 
 $et. al.$ \cite{PHYSICAL REVIEW B 82} through study on the thermal expansion coefficient and bulk modulus of Ni and Fe.

In similar line an attempt to study the effect of correlations by \textit{Łażewski} 
$et. al.$ \cite{PHYSICAL REVIEW B 74} present the effect of local Coulomb interaction \textit{U} on lattice dynamics in Fe including phonon density of states, equilibrium lattice constant and phonon frequencies, using the GGA+U method and report an upper bound for the value of effective U. \\
Seemingly, the literature lacks consensus on the stretch to which the influence of electronic correlations and magnetization to be held important in the lattice dynamics studies of correlated materials. The individual attempts to study either of the aspects within different theoretical frameworks i.e. DFT, DFPT has been seen to create more arbitrariness in the interpretation of results, as evident from few aforementioned studies which rather regard the findings to have huge exchange-correlation functional dependence than the inadequate account of correlation effects, suggesting implementation of combination of approximations. 
Notably, the available studies lack simultaneous account of correlation effects and magnetism to mark a comprehensive take on the estimations of phononic and related thermal properties per se.
The conclusions, drawn from the study of selected properties, could be viewed as property-specific and not generalizations to the material as a whole. For instance few of the above studies put remark on the effect of magnetism while studying the phonon frequencies along certain high-symmetric k-directions which might not bring about the true picture to disregard or regard the absolute effect, for other related lattice dynamic properties. Furthermore, the shortfall of substantial consistent efforts towards establishment of promising theoretical approach in this direction for correlated systems as "simple" as elemental transition metals raise the necessity for the course of research. We note that not much has been said about Ni metal in this regards ( on collective effect of correlations and magnetization) and being a correlated magnetic system, it serves the purpose of such study.
\\
In this direction, this work aims to address the arbitrariness by studying collective explicit dependence of correlation effects through on-site Coulomb interaction parameter $U$ and magnetism by analysis of difference in estimates obtained in both NM and FM phases of Ni. Consequently, in order to account for the experimental lattice dynamic properties of Ni, an attempt to seek for suitable theoretical approach is presented through a comparative DFT and DFT+U study. Apparantly, the bench-marking attempts for the $U$ parameter, by \textit{Sihi} \textit{et. al.} \cite{AntikV,AntikFe} report the cRPA calculated $U_{full}$ (wherein overall (effective/total) screening is constituted by including 3$d$ transitions along with other transitions in solid), commonly referred to as fully screened coulomb interaction parameter as the relevant choice of $U_{eff}$ to be used in DFT+U study. In this work the ($U_{full}$=0.516 eV) obtained from cRPA \cite{shivani}
 is used while carrying out DFT+U calculations. The FM phase of Ni corresponding to 0$K$ magnetization available in both the DFT and DFT+U formalisms is dealt with to understand the results pertaining to magnetic effects opposed to NM phase. Here, our results indicate prominent role of both correlations and magnetism in the estimation of phonon frequencies of Ni, visibly clear from phonon dispersion spectra and phonon density of states calculations. Similarly, the reflection of interplay of the effects is also found in estimates of derived thermal properties $i.e.$ phononic free energy, specific heat and entropy.  In addition to this, temperature dependent  phonon density of states (at 410 \& 940$K$) and thermal expansion coefficient are calculated for their comparative study and further equilibrium elastic constants and compressibilty factor are evaluated using both the techniques $i.e.$ DFT and DFT+U.  
\\
\section{Computational Details}
In this work, the electronic structure calculations are carried out for Ni, wherein full-potential linearized-augmented plane-wave method is used to carry out the non-magnetic (NM) and ferromagnetic (FM) calculations. The volume-optimized lattice parameter value of 3.513 Å is used with space group of 225 . Here, DFT and DFT+U calculations with PBE exchange-functional\cite{PBE} are carried out using WIEN2k code \cite{wien2k}. Phonon properties are calculated using PHONOPY code \cite{phonopy} based on finite displacement method (FDM) and supercell approach \cite{fdm}.
A supercell of size 2 × 2 × 2 is used for calculating the total forces on each atom in WIEN2k code. The k-mesh size of 5 × 5 × 5 is used in the full-BZ for force calculation. The convergence criterion for force calculations is set to 0.01 mRy/bohr. The forces are then used for calculating second-order force constants extracted using PHONOPY code to calculate the phonon frequencies. Thermal expansion coeffcient and related phononic thermal properties of Ni are also calculated under QHA as implemented in PHONOPY code.
\\
The process of calculating the phonon frequencies and finite-temperature phononic properties in this work, briefly to say includes calculation of force/atom using finite displacement method also referred to as  "frozen phonon" approximation on the supercell generated corresponding to the optimized lattice structure of Ni. The method involves displacement of an atom from its symmetrically allowed position in the generated supercell and thus the force on all other atoms is calculated which is numerically available from the converged self-consistent WIEN2k calculations.
The force/atom thus calculated is further used to calculate force constants and resultant dynamical matrix which upon diagonalization yeilds phonon modes or phonon frequencies using PHONOPY.
The equations below provide the resulting computed expressions for Force constant $\Phi$ and dynamical matrix $D$ as the function of atomic positions.
\begin{align*}
   F_{\alpha}(jl) = -\frac{\delta V}{\delta r_{\alpha}(jl)}
\end{align*}

\begin{align*}
   \Phi_{\alpha\beta}(jl, j^{'}l^{'}) = \frac{\delta^{2}V}{\delta r_{\alpha}(jl)\delta r_{\beta}(j^{'}l^{'})} = -\frac{\delta F_{\beta}(j^{'}l^{'})}{\delta r_{\alpha}(jl)}
\end{align*}
  Where $V$ stands for $V(r(j_{1}l_{1})......r(j_{n}l_{N}))$ which is to be inferred as general potential energy expression of a phonon system.
  where $r(jl)$ is the position of the $j^{th}$ atom in the $l^{th}$ unit cell with total $n$ atoms in a unit cell and the total no. of unit cells being $N$.
Further dynamical matrix is evaluated as-
 \begin{align*}
    D_{\alpha\beta}(jj^{i},q)= \frac{\Sigma_{l^{'}}\exp(iq.[r(j^{'}l^{'})-r(j0)])}{\sqrt{m_{j}m_{j^{'}}}}
 \end{align*}
further, reduces to eigenvalue problem-
  \begin{align*}
    \Sigma_{j^{'}\beta}  D_{\alpha\beta}(jj^{i},q) e_{\beta(j^{'},q\nu)} = [\omega(q\nu)]^{2} e_{\alpha(j,q\nu)}
  \end{align*}
 where m - atomic mass, q - wave vector, $\nu$ - band index, $e$ are the eigenvectors for corresponding band index and wave vector q, obatined by diagonalization of $D(q)$. The related thermodynamic properties are computed using the theoretical expressions provided below- 
  
  Harmonic phonon energy -\\
  \begin{align*}
    E = \Sigma_{q\nu} \hbar\omega(q\nu) [\frac{1}{2}+ \frac{1}{\exp(\hbar\omega(q\nu)/k_{B}T)-1}]
  \end{align*}
  
  Specific heat -\\
  \begin{align*}
    C_{V} = (\frac{\delta E}{\delta T})_{V}
          = \Sigma_{q\nu}k_{B} [\frac{\hbar\omega(q\nu)}{k_{B}T}]^{2} \frac{exp(\hbar\omega(q\nu)/k_{B}T)}{[exp(\hbar\omega(q\nu)/k_{B}T)-1]^{2}}
  \end{align*}
  
  Partition function -\\
 \begin{align*}
   Z = \exp(-\phi/k_{B}T) \Pi_{q\nu}\frac{\exp(-\hbar\omega(q\nu)/2k_{B}T)}{1-\exp(-\hbar\omega(q\nu)/k_{B}T)}
 \end{align*}

 Helmhotz Free energy -\\
 \begin{align*}
   F= -k_{B}T\ln Z 
 \end{align*}
\begin{align*}
    F= \frac{1}{2}\Sigma_{q\nu}\hbar\omega(q\nu) + k_{B}T\Sigma_{q\nu}\ln [1-\exp(-\hbar\omega(q\nu)/k_{B}T)
\end{align*}  

Entropy -\\
\begin{align*}
  S = -\frac{\delta F}{\delta T}
\end{align*} 
\begin{align*}
 S = \frac{1}{2T}\Sigma_{q\nu}\hbar\omega(q\nu)\coth(\hbar\omega(q\nu)/2k_{B}T)\\
-k_{B}\Sigma_{q\nu} \ln[2\sinh(\hbar\omega(q\nu)/2k_{B}T)]
\end{align*}

The elastic constants are calculated using long wave method as provided by \textit{Barker} \textit{et al.} \cite{PhysRevB.2.4176}

Wave velocities $v$ are calculated for the longitudinal and transverse waves in the $<100>$ and $<110>$ directions using the equation -

\begin{align*}
  v = \frac{\delta \nu}{\delta q}
\end{align*}

Where $\nu$ is phonon frequency and $q$ the phonon wave vector, evaluated at $q$ = 0.
The elastic constants $c_{ij}$ are calculated as:

\begin{align*}
  c_{11} = \rho  v_{1}^{2} \\
  c_{44} = \rho  v_{2}^{2} 
\end{align*}

\begin{align*}
c_{12} = c_{11} - 2\rho  v_{3}^{2}
\end{align*}

where $v_{1}$,$v_{2}$,$v_{3}$ are wave velocities along longitudional $<100>$, transverse $<100>$ \& slow transverse $<110>$ waves.\\
Compressibilty factor is found as -

\begin{align*}
  K = \frac{3}{c_{11}+2 c_{12}} 
\end{align*}

\section{Results and Discussion}
To study the effect of electronic correlations and magnetization on the lattice dynamics of Ni, the calculations have been carried out at DFT and DFT+U level in both NM and FM phases. In order to see the effect, phonon dispersion and phonon density of states (DOS) have been computed and compared with the available experimental results.
Fig.1 shows the calculated phonon dispersion curves with the room temperature experimental curve\cite{expt_band} obtained along $\Gamma$ to X and $\Gamma$ to L high symmetric k-directions. It contains three acoustic (one longitudinal and two-degenrate transverse) branches.
\begin{figure*}
    \centering
    \includegraphics[width=13cm]{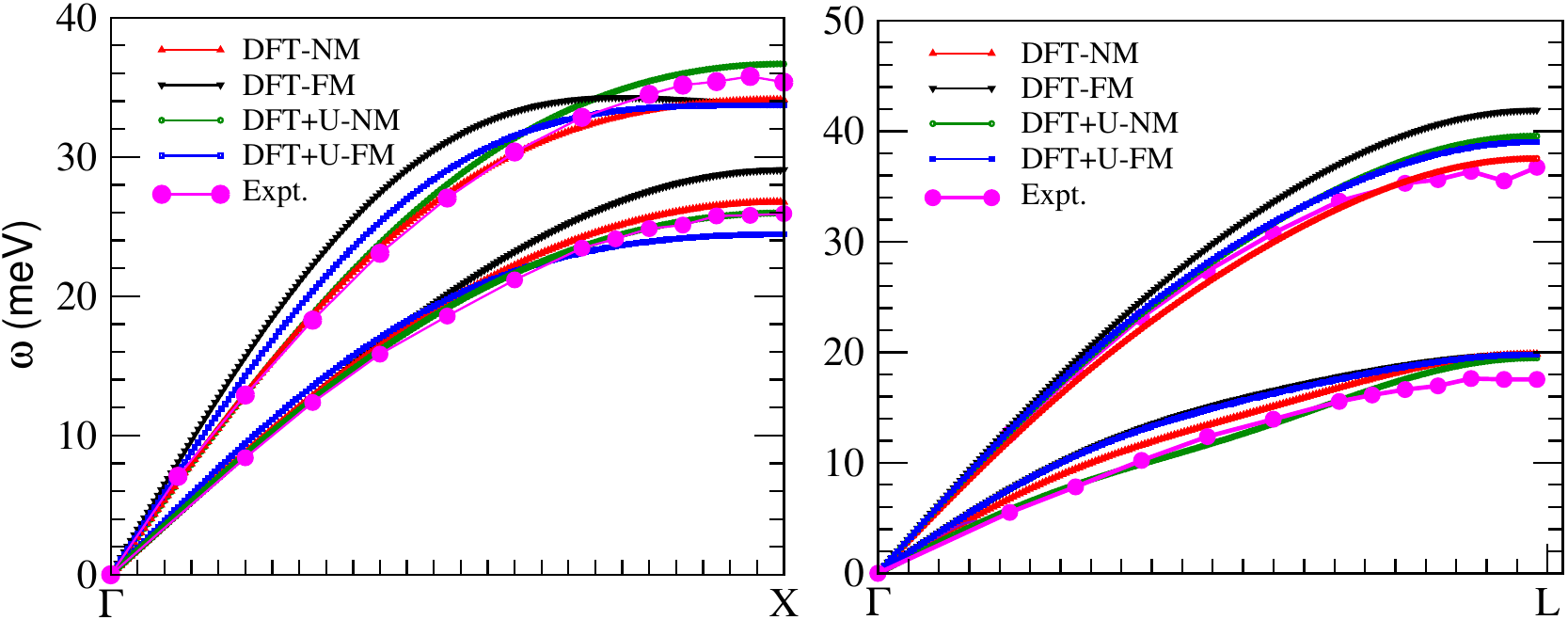}
    \caption{Calculated phonon dispersion spectra along $\Gamma$ to X \& $\Gamma$ to L, high symmetric k-directions using DFT and DFT+U, in NM and FM phases of Ni, along with experimental curve \cite{expt_band} available at room temperature}.

    \label{fig:}
   \end{figure*}
The evident difference in the curves belonging to NM and FM phases, at both the DFT and DFT+U levels, suggests the influence of magnetization on the phonon frequencies along the directions. The frequency values obtained in two phases differ more in case of longitudinal branch than transverse. At DFT (DFT+U) level, maximum frequency deviation of about 4$meV$ ( 2$meV$) corresponding to the two phases has been found in the midway region (towards X point) in $\Gamma$ to X direction and ( towards L point) in $\Gamma$ to L direction.
Calculated phonon dispersion curves indicate the effect of electronic correlations and magnetization on the estimation of phonon frequencies. It becomes important to note that within the experimental error bar (around 0.5$meV$) DFT+U-FM and DFT+U-NM both can be regarded in good agreement with the room temperature experimental data for both the branches.

Further, for having closely whole picture of full-BZ, phonon DOS has been calculated. Fig.2 shows the calculated phonon DOS obtained from the same set of formulations subjected to the broadening of around 1.13$meV$ along with the available experimental data at 10$K$ \cite{expt_dos} . The characteristic features visible in experimental data include peaks A and B at frequencies 33$meV$ and 24$meV$, respectively and a hump like feature C at around 22$meV$, along with a dip D at 30$meV$. The data show highest phonon frequency cut off at around 40$meV$, which is seen to be appreciably accounted by DFT-NM curve. The DFT curves of both the phases can be seen to account for the peak A but fail to account for the experimental features B, C and D. Whereas, DFT+U in NM phase improves on to account for the dip as well along with peak A while gets more worse around the frequency region of the features B and C. The DFT+U-FM curve comes out to be in overall good agreement with the experimental curve accounting for the visible features as well as qualitative behavior in both the low and high frequency limits.

\begin{figure}
    \centering
    \includegraphics[width=6cm]{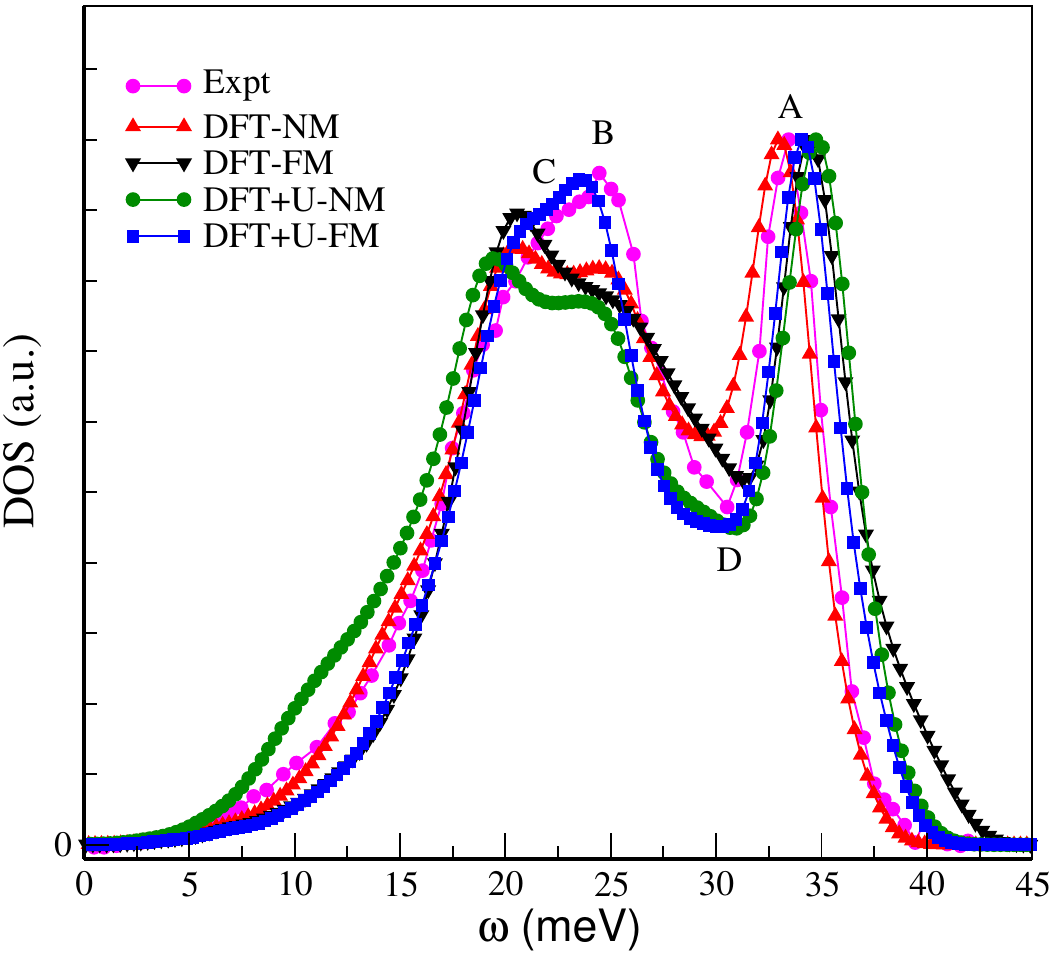}
    \caption{Calculated Phonon DOS using DFT and DFT+U in NM and FM phases of Ni, along with 10$K$ experimental data curve. \cite{expt_dos}.}
\end{figure}

As mentioned before 
the study \cite{PhysRevB.62.273} done on Fe and Ni within DFPT framework along high-symmetric k-directions in magnetic phases, shows weak role of magnetization on phonon frequencies. In contrast to this study we find significant effect of magnetization on frequencies along with electronic correlations, which also becomes evident from phonon DOS plot. Other such study by Lee $et$ $al.$ \cite{Lee} 
using DFPT at fixed magnetic moment (FSM) employing LDA functional, suggest the importance of consideration of magnetization in calculating phonon frequencies meanwhile showing an opposite trend between increasing magnetic moments to what our study reflects for NM and FM phase, notably their FSM calculations depict still overestimation of experimental frequencies. Yet another attempt of study on Fe  \cite{PHYSICAL REVIEW B 74}
suggests the influence of Coulomb interaction parameters in calculating the phonon frequencies, which is also evident from our study.

\begin{figure*}
    \centering
    \includegraphics[width=12cm]{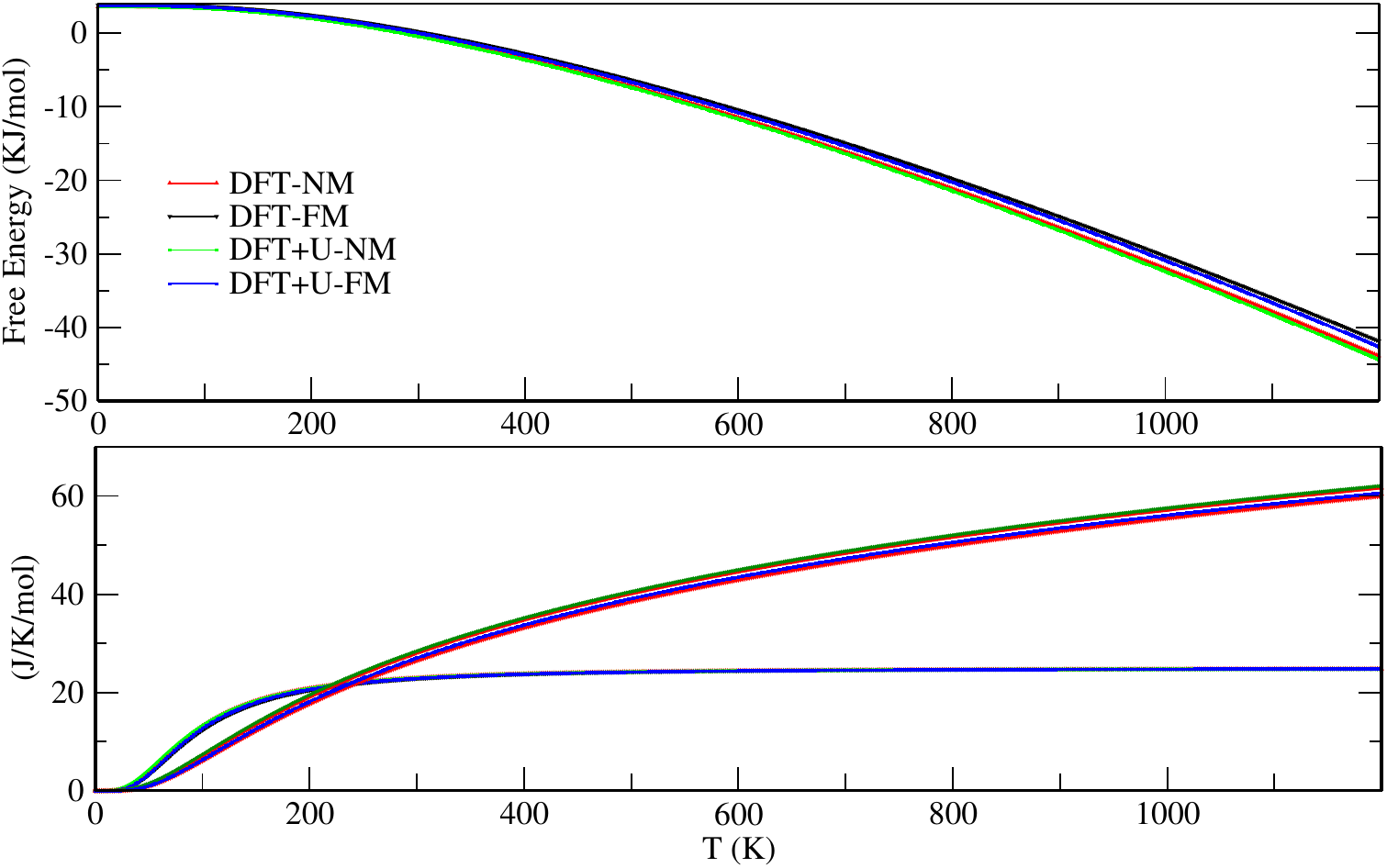}
    \caption{ Calculated thermodynamic properties' variation with temperature in Ni. Above: Helmholtz free-energy Below: specific heat and entropy. }
\end{figure*}
From the above results where, DFT+U-FM considerably gives realistic estimation of phonon frequencies at low temperatures, it could be expected to account for the lattice dynamic properties as well, which follow from phonon frequencies $i.e.$ thermal properties especially in low temperature region.  
Further part of work involves the calculations of thermal properties pertaining to the lattice dynamics- free energy, entropy and specific heat.
The phononic free energy given in Fig.3 shows relatively negligible difference in calculated curves in low temperature region ranging from 0 - 200$K$. The curves start to deviate from each other towards high temperature range $i.e.$ at 600$K$, DFT-NM and DFT+U-NM lie close to each other (at $\sim$ -11.4 $KJ/mol$) showing less significant effect of $U$ in NM phase, Whereas, comparatively in FM phases DFT-FM and DFT+U-FM, curves lie apart by nearly 0.4$KJ/mol$ at 600$K$. The curves in FM phase differ slightly more around 1200$K$ region by $\sim$1$KJ/mol$, while the NM phase curves differ by  $\sim$0.5 $KJ/mol$. It could also be noted that free energy is estimated at relatively higher values by the FM phase in comparison to their respective NM phase. 
The calculated entropy curves deviate from each other with seemingly constant difference in both the NM and FM phases in the temperature range of 500$K$ to 1200$K$. Both the NM curves maintain a difference of around 0.2$J/K mol$, whereas the FM curves lie slightly below the NM curves, estimating entropy by 0.4$J/K mol$ difference.
Further, the specific heat calculations suggest no significant difference in the calculated curves for the phononic energy in high temperature region ranging from 500-1200$K$. The constant values achieved by the curves around 500$K$ correspond roughly to the excitation of highest phonon mode with frequency ($\sim$40$meV$) whereafter further significant contribution to the specific heat ceases ( consistent with the Dulong-petit's law). The curves differ in low temperature region $\sim$200$K$ with DFT-FM and DFT+U-NM showing maximum deviation among all curves by $\sim$ 1.5$J/K mol$.

In line of studying the thermal properties, thermal expansion coefficient $\alpha$ values over a temperature range of 0-1000$K$ have been calculated within the above mentioned frameworks, as given in Fig.4. In the low temperature region roughly till 150$K$ all the curves show finely close estimates of values, while remarkable difference in values (by maximum of $\sim$$0.25$$\times$$10^{-6}$ units) is found between DFT+U-NM and DFT+U-FM in the temperature region around 200$K$. In comaparison to DFT+U curves, DFT curves in both the phases are seen to consistently overestimate, in the region roughly ranging from 200$K$-400$K$. In temperature regime from 600$K$ to 1000$K$, DFT+U-NM curve can be observed to exhibit notable change in behaviour around 600$K$, followed by a dramatic rise in values. Comparative analysis suggests agreeable correpondence of DFT+U-FM curve, together with the DFT-NM and DFT-FM curves with the experimental results in the temperature range extending up to 300$K$. The 300$K$ estimates are found to be $1.2$$\times$$10^{-5}$, $1.19$$\times$$10^{-5}$ and $1.11$$\times$$10^{-5}$ by DFT-FM, DFT-NM and DFT+U-FM respectively, where the experimental\cite{expt_alpha} room temperature value lies at around $1.28$$\times$$10^{-5}$ units. The failure of DFT+U-FM above 600$K$ and subsequently the behaviour of DFT+U-NM can be recognized in qualitative respects given that Ni exists in the FM state below 631$K$. Additionally, at high temperatures, the increased deviation in calculated and experimental $\alpha$ values, does not seem surprising as here QHA has been used, where the electronic ground state energies are taken corresponding to different volumes, while calculating $\alpha$.

\begin{figure}
    \centering
    \includegraphics[width=6cm]{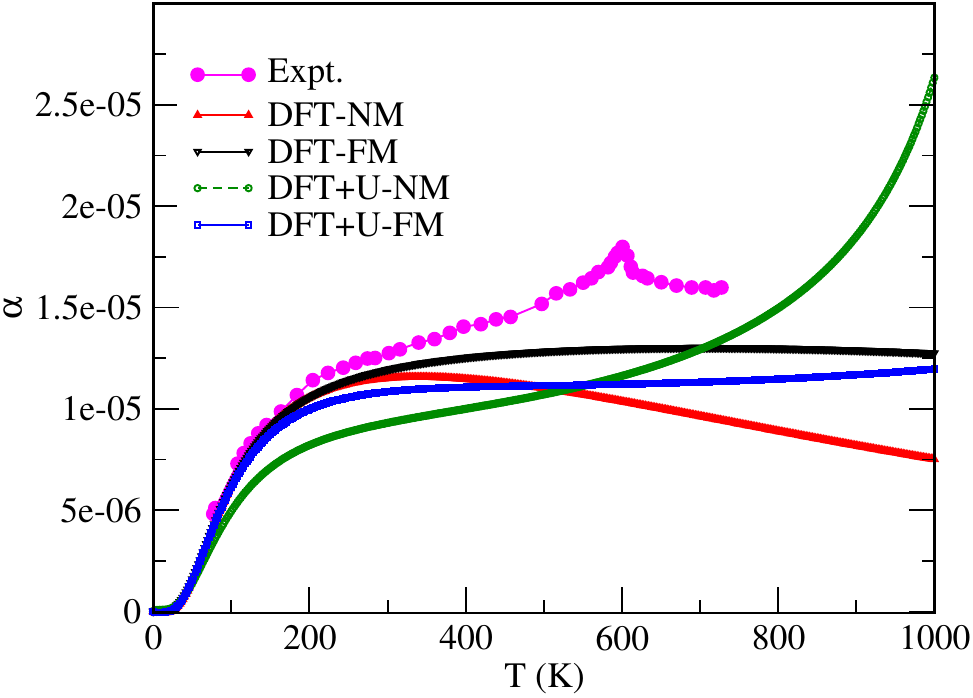}
    \caption{ Calculated temperature-dependent thermal expansion coefficient ($\alpha$) using DFT and DFT+U in NM and FM phases of Ni, along with experimental curve \cite{expt_alpha} at room temperature.}
\end{figure}

\begin{figure}
    \centering
    \includegraphics[width=6cm]{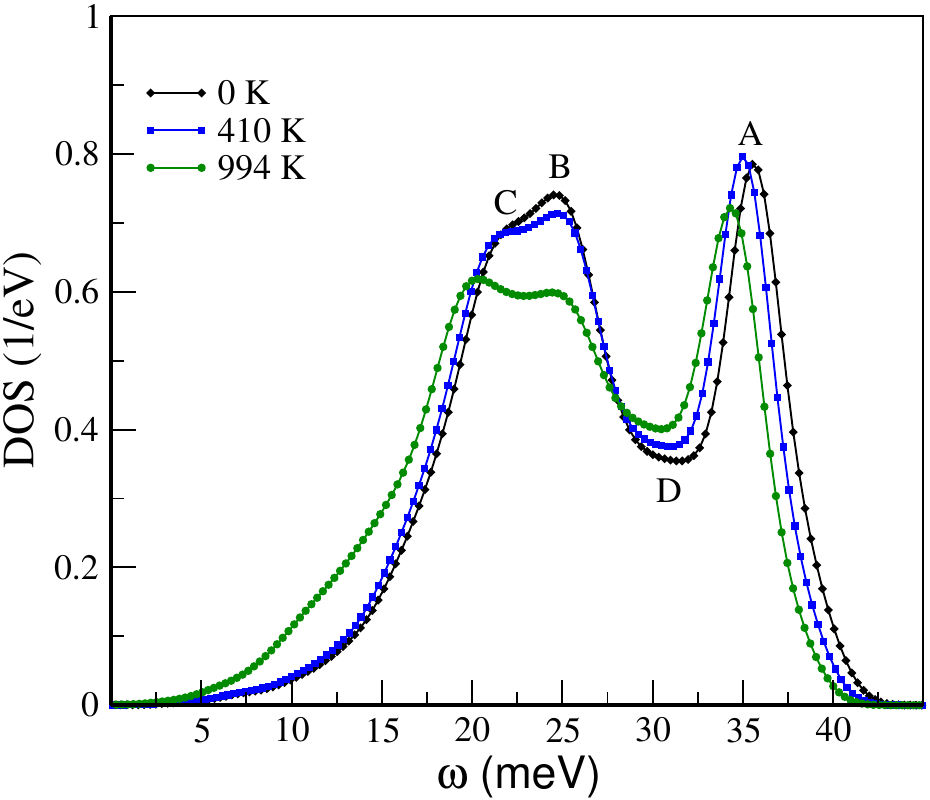}
    \caption{Calculated phonon DOS at 0$K$, 410$K$ and 940$K$ for Ni.}
\end{figure}

Further course of this work deals with investigating the thermal effects on the phonon DOS. Since, as already discussed DFT+U-FM provides better description of experimental phonon DOS at 10$K$, and to an extent DFT+U-NM is able to produce its charachterstic features as well, the figure here includes curves corresponding to two finite temperatures $i.e.$ 410 and 994$K$ pertaining to $1\%$ and $2\%$ percent volume expansion, respectively within DFT+U framework in both the phases based on temperature value below or above $T_c$, along with 0$K$ curve. The figure clearly indicates the effect of temperature showing the shift in peaks A and B towards low frequency region, as we go from 0- 410 - 994$K$, togetherwith resulting in significant broadening as reflected from the reduced intensity of peak especially visible around B peak. The calculated curves depict the effect of temperature consistent with the experimental observtions provided by Delaire $et$ $al.$\cite{expt_dos} \\

 \begin{table}[h]
    \centering 
    \caption{ Elastic constants ($C_{11}$, $C_{12}$, $C_{44}$) (in the units of $dyne/cm^{2}$) and compressibility (K) of Ni ((in the units of $cm^{2}/dyne$) )}
     \begin{tabular}{c|c|c|c|c}
      \hline
      Method & $C_{11}(10^{12})$ & $C_{12}(10^{12})$ & $C_{44}(10^{12})$) & $K(10^{-12})$ \\
      \hline
      DFT-NM & $3.32$ &$0.69$ &	$1.15$  &	$0.637$ \\
      DFT-FM & $4.00$ & $1.50$ &	$1.42$ & $0.427$	\\
      DFT+U-NM & $3.77$ & $1.23$ & $0.83$ & $0.480$ \\
      DFT+U-FM & $3.85$ & $0.75$ & $1.41$ & $0.558$ \\

     Expt\cite{neighbours} & $2.52$ & $1.52$ & $1.23$ & $0.538$ \\
     Theory \cite{theory_elastic}      & $2.4$ & $1.4$ & $1.4$ & $0.576$\\
      \hline
      \end{tabular}
    
    \label{tab:}
  \end{table}
\vspace{1.0cm}

Finally the elastic constants along with compressibility factor have also been calculated and provided in the table along with available room temperature experimental\cite{neighbours} and theoretical\cite{theory_elastic} results. Evidently, DFT in FM phase is found to estimate elastic constants at relatively higher values than other calculated estimates. Seemingly, the methods tend to overestimate the elastic constants when compared to the available experimental and theoretical study by almost $50\%$ in case of DFT+U-FM $c_{11}$'s estimation. The compressibilty factor resulting from DFT+U-FM ($0.558\times 10^{-12}$) seems to be in close agreement with experimental and theoretical findings.  The obtained calculated values could be considered to an extent reasonably well while regarding deviation to the prominence of thermal effects taking over at finite temperature.

 \section{Conclusion}
 We present a comprehensive study of phonon and related phononic thermal properties for their electronic correlations and magnetic phase dependence through a step-by-step account of both the aspects realized via a comparative DFT and DFT+U study on Ni. The calculations carried out for the phonon-dispersion spectra and phonon density of states suggest considerable difference in obtained frequency values at the DFT and DFT+U level for both the NM and FM phases, thus indicating the effect of both the aspects. We find DFT+U-FM estimates appreciably close to the experimental frequencies suggesting the role of both the aspects ( $U$ and magnetic ordering). The relative success of DFT+U in FM phase to account for the realistic description of experimental phonon frequencies in Full-BZ (which being a broader picture of k-dependent properties' evaluation) suggest the importance of simultaneous account of correlations and magnetic ordering in Ni.
 Notable deviation in all the four formulations (DFT-NM \& FM, DFT+U-NM \& FM) in case of thermodynamic properties again depicts the role of both the effects, specially towards high-temperature region.
  Conclusively, we establish the prominence of effects further after studying the equilibrium elastic constants and compressibility factor and the temperature dependent phonon density of states \& thermal expansion coefficient.



\end{document}